# Changes in Ugandan Climate Rainfall at the Village and Forest Level

[1]Paddy Ssentongo, [3]Abraham J. B. Muwanguzi, [4]Uri Eden, [5]Timothy Sauer,

[3]George Bwanga, [6]Geoffrey Kateregga, [7]Lawrence Aribo, [7]Moses Ojara,

[3]Wilberforce Kisamba Mugerwa, [1,2]Steven J. Schiff

[1]Center for Neural Engineering, Department of Engineering Science and Mechanics,
[2]Departments of Neurosurgery and Physics, The Pennsylvania State University, University Park, PA

[3]Ugandan National Planning Authority, Kampala, Uganda

[4]Department of Statistics, Boston University, Boston, USA

[5]Department of Mathematics, George Mason University, Fairfax, VA, USA

[6]Map Uganda, Kampala, Uganda

[7]Ugandan National Meteorological Authority, Kampala, Uganda

Correspondence:

Steven J. Schiff
Center for Neural Engineering
The Pennsylvania State University
University Park, PA 16802 USA
steven.j.schiff@gmail.com

## Abstract

In 2013, the US National Oceanographic and Atmospheric Agency refined the historical rainfall estimates over the African Continent and produced the African Rainfall Climate version 2.0 (ARC2) estimator. ARC2 offers a nearly complete record of daily rainfall estimates since 1983 at 0.1 x 0.1 degree resolution. Despite short-term anomalies in twice-yearly rainy season intensities in Uganda, we identify an overall decrease in average rainfall of about 12% during the past 34 years. Spatiotemporally, these decreases are greatest in agricultural regions of central and western Uganda, but are also reflective of rainfall decreases in the gorilla habitat within the Bwindi Forest in Southwest Uganda. The findings carry significant implications for agriculture, food security, and wildlife habitat.

**Africa is both the driest and hottest of continents, and its available water is essential for almost all human activities and to support non-human biodiversity.[1] In 2013, the US National Oceanographic and Atmospheric Agency refined the historical satellite-based rainfall estimates over the African Continent and produced the African Rainfall Climate version 2.0 (ARC2) estimator[2]. This estimator combines daily geostationary rainfall estimation through infrared cloud reflectivity with ground based rainfall measurements at a fine grid scale at 0.1° x 0.1° resolution (approximately 11x11 km at the Equator). The record reaches back to 1983, and continues with real-time daily data production. We here report an analysis of the stationarity and climate patterns of rainfall affecting Uganda over the period 1983-2016.**

There are many critical uses of such data at this fine a scale. From a human planning perspective, major construction projects can be mapped to the mean and variability of rainfall for a given location or extent, and engineered to account for anticipated extreme events. Additionally, agricultural planning can be adapted to patterns and predicted changes in rainfall. Moreover, since in a country like Uganda about 70% of the population depends on agriculture, these data may be used to understand population growth and density in given regions and therefore facilitate planning for settlements and economic activities. From a non-human perspective, habitat limitations for endangered species may place their survival at greater risk.

Many human infectious diseases have strong relationships to rainfall: cholera[3], malaria[4], leptospirosis[5], melioidosis[6], and the seasonal Neisseria meningitis within the African meningitis belt[7]. Infant infections leading to postinfectious hydrocephalus have been noted to have a significant relationship to rainfall[8].

Geographically, Uganda stretches from 1.43°S to 4.27°N and from 29.5°E to 35.03°E. At 0.1° x 0.1° resolution Uganda is contained within a square 61x61 grid. In Figure 1A we illustrate the country as a composite of the boundaries of the 44,034 villages that comprise the landmass, and superimpose the 3,721 satellite grids overlay.

The entire data set for all 3,721 time series (plotted in sequential colors) is illustrated in Figure 1B. The cumulative rainfall over 34 years is mapped onto the satellite grid in Figure 1C. Fusing these with the country map in Figure 1D, there are several notable features. In the northeast, the semi-arid region of Karamoja is shown with low cumulative rainfall. The heaviest rainfall is in the northwest over the Congo River basin rainforest. The other region with high rainfall is where the northeast edge of Lake Victoria meets the Ugandan landmass in the lower central region of the plot. The averaged power spectral density of all 3,721 time series with error bounds is shown in Figure 1E, which reflects the dominant 1 and 2 cycle per year frequencies, and the spectrogram in Figure 1F demonstrates the consistency of these two fundamental frequencies throughout the 34 year record. The 2-cycle per year rainfalls in the East African Highlands are of unequal size, augmenting the 1-cycle per year frequency amplitude.

Statistically, rainfall distributions are often non-Gaussian due to non-negativity and pronounced skewness. For such data, the normal distribution will inadequately account for the rainfall variability. This renders ordinary least squares, with the assumed normal distribution of errors, a problematic choice for model fitting. Indeed, the rainfall distribution for all 46,211,099 daily measurements is highly skewed (Figure 2A), and it is well known that such rainfall data may follow a gamma distribution[9]. By averaging across the 3,721 spatial locations for each day in time, and filtering between frequencies of 1/20 to 6 cycles per year for visualization, the twice-yearly rainy season cycles are now readily visualized (Figure 2B), and the distribution becomes much less skewed for the 12,419 days of data (Figure 2C). These data can all be appropriately fit using the generalized exponential family of distributions modeled within the Generalized Linear Model (GLM) framework that embraces such distributions ranging from gamma to normal.[10]

In Figure 2D we demonstrate a GLM fit to the 34 years of data, spatially averaged for each day across the spatial grid but not filtered over time. We show both a log-linked linear fit with time ($\log(\mu)=A+BT$, where $\mu$ is the expected rainfall), as well as a fit of both time and a linear combination of frequencies (using 4, 2, 1, 0.5, and 0.25 cycles per year). There is a significant downward slope of the GLM dependency on time of the spatially averaged rainfall by 12% over the 34-year interval (slope of -0.0038 corresponding to an exp(-.0038)=0.38% reduction in rainfall per year, $p<1.6 \times 10^{-5}$, slope standard error SE = 0.00088), and the quality of the fit of time and frequencies is also highly significant (F vs constant model 146, $p < 7 \times 10^{-317}$).

To examine the spatial contributions to this decrease in rainfall, we fit 3,721 GLM models to each spatial location's time series (without averaging or filtering). The origin of the average negative slope, B, for the linear model fit, $\log(\mu)=A+BT$ in Figure 2D, is based upon a distribution of slopes with a mean of -0.0031 (0.31% reduction per year), where the probability (fraction) of negative slopes is 0.78 (2,902/3,721). We now plot these individual slopes in their spatial location on the satellite grid in Figure 2E. The color map for the spatial distribution of these slopes uses the intensity of brown and green to represent the magnitude of the decrease or increase in slope respectively. There is broad region in central and western Uganda that appears to be responsible for the overall decrease in rainfall shown in Figure 2D.

These 3,721 slopes (Figure 2E) represent a massive multiple testing problem. To explore this spatial distribution further, we turn to the control of false discovery rate (FDR) using the method of Benjamini and Hochberg.[11] We plot the curve representing the GLM goodness of fit as p-values (in blue) in Figure 2F against a family of FDRs ranging from 0.02 to 0.2, along with the standard FDR for a single test (0.05) and the Bonferroni corrected false positive rate of $1.3 \times 10^{-5}$. The distributions corresponding to these FDRs are shown in Figure 3A, and their corresponding spatial maps in Figure 3B. As the false discovery rates vary from 0.1 to the Bonferroni rate, identification of the region of central and western Uganda with decreasing rainfall over these 34 years remains robust.

We can independently test these decreases in rainfall by taking difference maps of cumulative rainfall for different periods of time. In Figure 3C, we illustrate the difference of cumulative rainfall in the first vs most recent 15, 10 and 5 years of the data, along with the second 5 vs next to last 5 year segments of data. The spatial maps illustrating increases or decreases in rainfall illustrating a broad region of decreased rainfall within central Uganda remain reflective of the GLM regression slopes in Figure 3A and 3B.

Next, we examine the Bwindi Impenetrable Forest within Uganda. This Forest reserve constitutes one of the last remaining habitats of the Mountain Gorilla, and is considered crucial for species survival. Ground based rainfall data is understandably incomplete[12]. The coordinates of the 331 square km Bwindi Forest lies within 0.85°S and -1.15°S, and 29.55°E and 29.85°E on the satellite map. These coordinates correspond to 16 squares of our grid (Figure 4A). The cumulative rainfall from ARC2 is shown in Figure 4B. We again find that there is a significant downward slope of the GLM dependency on time of the spatially averaged rainfall by 12.8% over all years (slope of -0.004 corresponding to a 0.4% decrease in rainfall per year, $p<0.01$, slope SE = 0.00016), consistent with our findings in the entire grid.

Lastly, both the El Niño Southern Oscillation (ENSO) and the Indian Ocean Dipole[13] (IOD) are known to influence rainfall in East Africa.[14] We turn to the technique of wavelet coherency, and seek a statistical bootstrap that preserves more of the properties of the data than the white[15] or colored[16] noise employed in previous work on geophysical systems. Following the recommendation for a non-parametric bootstrap constructed from surrogate data from the original time series[17], we used a randomization scheme previously employed by swapping binary partitions at random locations in a time series,[18] to ensure that local correlations are effectively destroyed across an ensemble of such resampled time series, yet the relevant frequencies and distribution of values remain the same. Applying this bootstrapped statistical method, there are regions at the 99% confidence limit in the latter half of our time series where ENSO (Figure 5A, 2006-2014, 1-2 year periods) and IOD (Figure 5B, 2004-2015, 1-4 year periods) coherency with rainfall remains significant.

The relatively short length of the instrumented ARC2 limits our analysis to 34 years. Proxy and model simulation suggests that the IOD is more important than ENSO over multidecadal and perhaps longer time scales.[19] Although ENSO effects interannual droughts[14], it is less important for interdecadal rainfall patterns[20], and our results are consistent with this. Nevertheless, our findings that ENSO and IOD effects on Ugandan rainfall have been more significant within the more recent half of the 34-year record remains unexplained.

The weather patterns in East Africa are unusually complex and regionally disparate[21]. Uganda is at the western edge of the Greater Horn of Africa[22] (GHA) region, and has historically had more rainfall than neighboring Kenya and Tanzania[21]. Although many climate models predict that East Africa will experience an increase in rainfall as the planet's atmosphere warms, it has in fact become drier over recent decades[19] , and resolving this discrepancy has been a topic of active research.[20]

One of the hazards in generalizing from regional models to more localized regions, or extrapolating using proxy data taken from highly localized sediment core analysis[23], is that predictions may not equally apply to all countries within such regions. For instance, the multi-decadal influence on East African rainfall predicted from IOD dynamics[19] is not well reflected in our 34-year Ugandan dataset, where the effects appear sub-decadal.

How such rainfall decreases impact infectious disease prevalence and risks will be determined by individual disease characteristics, and will be important for specific locations. By fusing the satellite rainfall grid with the locations of all of the villages in Uganda, we have a finely granular way to track epidemic disease. Such fusion is a platform for seeking optimization of treatment, and prevention, of many infectious diseases.

Although climate is global and regional, policy and preparation remains largely dependent upon individual countries. Uganda is a country where 72% of geographical area is used for rain-fed farming and the population growth is one of the highest in the world.[12] An average rainfall decrease of this magnitude, over the multiple decades of the climate record examined here, is important for agriculture in a country dependent on subsistence crop yields. There is a substantial need for more granular and accurate prediction modeling for both short-term drought anticipation and longer-term rainfall trends within the time-frame relevant for economic planning. Nevertheless, the present trend in rainfall decrease is gradual enough so that there remains an opportunity to build adaptive capacity[1] through strategies[22] to make the country more resilient: anticipatory land-use management, shifts towards more sustainable agricultural practices, and infrastructure development to increase the resiliency of the society with respect to short and long-term changes in rainfall.

Acknowledgements:

We are grateful to Fredrick Kayanja (Gulu University), Mujuni Godfrey (UNMA), Tumusiime Moses, (UNMA), Godwin Ayesiga (UNMA), Otim F. C. Obeke (UNMA), Tenywa Joseph (NPA), Sajjabi Fredrick (NPA), Ongora Emmanuel (NPA), Namyalo Jackie (NPA), Arineitwe Justine (NPA), and Tebugulwa Allen (NPA), for their helpful discussions. Supported by US NIH Pioneer Award 5DP1HD086071.

## Methods

Data

Rainfall data was obtained from the gridded, daily 34 year precipitation estimation dataset (http://www.cpc.noaa.gov/products/international/data.shtml) centered over Africa at 0.1$^{\circ}$ x 0.1$^{\circ}$ spatial resolution, the African Rainfall Climatology, version 2 (ARC2)[2]. These data are an estimation derived from a fusion of the geostationary infrared sensing from the European Organisation for the Exploitation of Meteorological Satellites, and 24 hour rainfall measurements from Global Telecommunication System gauge observations. There are 341 missing days in these ARC2 data (out of 12,419 days) which were accounted for by linear interpolation.

El Nino Southern Oscillation data (ENSO) was obtained from NOAA at https://www.esrl.noaa.gov/psd/enso/mei/table.html. These monthly data through 2016 were expanded based on the length of each month into equivalent daily data.

Indian Ocean Dipole (IOD) data was obtained from The Extended Reconstructed Sea Surface Temperature (ERSST) dataset derived from the International Comprehensive Ocean–Atmosphere Dataset (ICOADS) at NOAA at https://iridl.ldeo.columbia.edu/SOURCES/.NOAA/.NCDC/.ERSST/.version4/.IOD/.C1961-2015/.iod/datatables.html. These monthly data through 2015 were expanded based on the length of each month into equivalent daily data.

The geocoded map of Uganda at the village level was compiled employing census and election commission records, and publically accessible at [url to be established at the Ugandan National Planning Authority with publication of this paper].

Signal Processing

Spectra were calculated, following removal of the mean from each of the 3,721 time series, using Hamming data windows 3-years in length (3*365 days), with 95% overlap of these windows, and 1-sided spectral estimations performed using Welch's method as implemented in Matlab function *pwelch*.[24] The power spectral density (PSD) from each of the 3,721 time series from the different locations were then averaged, and plotted on a decibel scale (10*log(PSD)) in Figure 1E for frequencies greater than zero and less than 12 cycles per year. The spectra from each windowed time period were then assembled into the spectrogram in Figure 1F.

When filtering was applied, such as in Figure 2B, we employed a bandwidth of 1/20 to 6 cycles per year, an order of 100, frequency of 365 days per year, Nyquist frequency of 365/2, and a finite impulse response (FIR) filter (Matlab function *fir1*) applied in a zero-phase distortion manner using Matlab function *filtfilt*. The spatial mean of all 3,721 filtered time series were plotted in Figure 2C.

When employed in wavelet coherency, both the spatially averaged rainfall time series (through 2015), and ENSO time series were equivalently filtered within the same 1/20 – 6 cycle per year bandwidth.

<u>Statistical Analysis</u>

*Generalized Linear Modeling*

We implemented generalized linear modeling (GLM) using the methods developed by Nelder and Wedderburn.[10,25] Using the implementation of the Matlab function *glmfit*, we employed gamma distributions and log link functions.

The false discovery rate (FDR) was controlled using the method of Benjamini and Hochberg.[11] A family of FDR rates was developed, assuming that the false positive (Type I) error rate was $\alpha$ (nominally 0.05), the number of comparisons tested was $m$, as the largest $p(i)$, where:

$$p(i) \leq \alpha * \frac{i}{m}, \qquad 0 \leq i \leq m.$$

These $p(i)$ thresholds, for a family of FDR rates, are shown as intersections between the blue and red lines in Figure 2F.

<u>Wavelet coherence</u>

We employed the method of Torrence and Webster[15], as implemented in the Matlab function *wcoherence*, except we replace their white noise surrogate with surrogates based upon randomly partitioned[18] and time reversed surrogates based upon one of the time series. We chose the partitioned sections of the randomly, swapping the partition order. This has proven a robust method to break up short term correlations between two time series, while preserving the statistical properties other than losing the slowest of frequencies due to the partitioning. Unlike preserving spectra by randomizing the phases of frequencies in a Fourier transform, and then inverting the transform, this method also preserves the original data values and distribution,[18] and is much more computationally efficient than simulated annealing alternatives.[26] We created an ensemble of 1000 such surrogate wavelet coherencies, chose the largest 95% or 99% surrogate coherencies at each point in the time-frequency plot, and used these thresholds to build our bootstrapped confidence limits setting all values to zero which did not exceed these thresholds. We plot the remaining significant wavelet magnitude-squared coherencies in the significance plots of Figure 5.

<u>Code availability</u>

The data sets, and Matlab code required to replicate the findings of this papers, are openly available at [url with durable link to be established at the time of publication of this paper].

## Figure 1

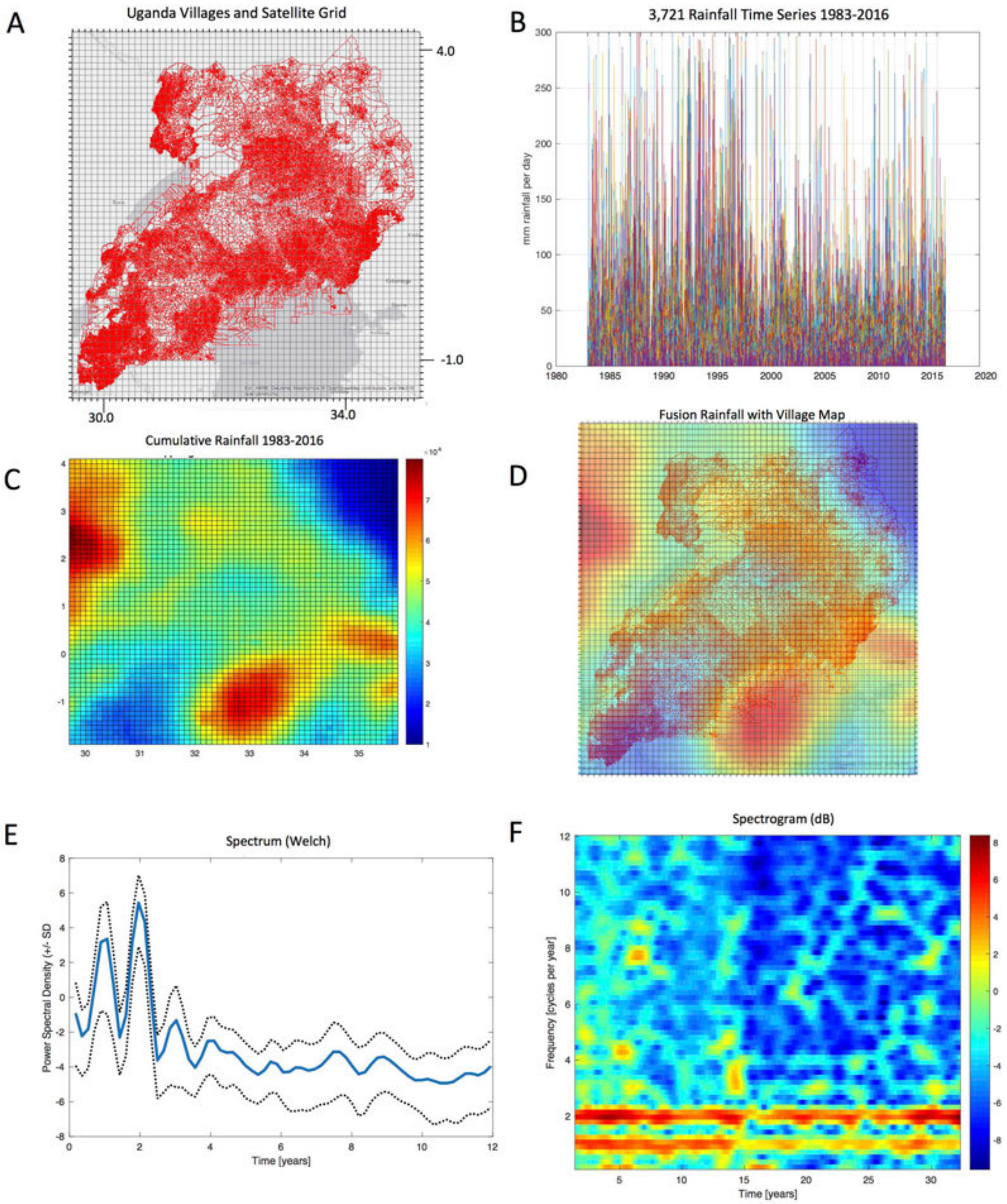

*Figure 1.* Fusion of satellite rainfall data with map of Uganda at the village level. Shown are the pattern of cumulative rainfall over the 34 year dataset, and the frequency content of the rainfall demonstrating twice-yearly rainy seasons. A) Satellite grid overlay over Uganda showing the 44,034 villages as polygons, overlain with the 61x61, 0.1°x0.1°, satellite grid. B) Rainfall per day from each of 3721 grid locations for 12,340 days. C) Cumulative rainfall on 61x61 grid for all 12,419 days over 34 years from 1983 – 2016. D) Fusion of cumulative rainfall with village and country map. Note that the heaviest rains are over the rainforest within the Congo River Basin to the west of Uganda, and where the northeast corner of Lake Victoria abuts the Ugandan land mass. The driest region is in in northeast Uganda, where the Karamoja district abuts northwest Kenya and South Sudan. E) Spectral density estimated from each of 3,721 grid locations over 12,419 days, with mean and ± 1 SD. Hanning window of 3x365 with 95% overlap. Points plotted as decibels ($10\log_{10}$). F) Spectrogram with mean removed from signal. Note the dominant rainy cycles at 1 and 2x per year.

**Figure 2**

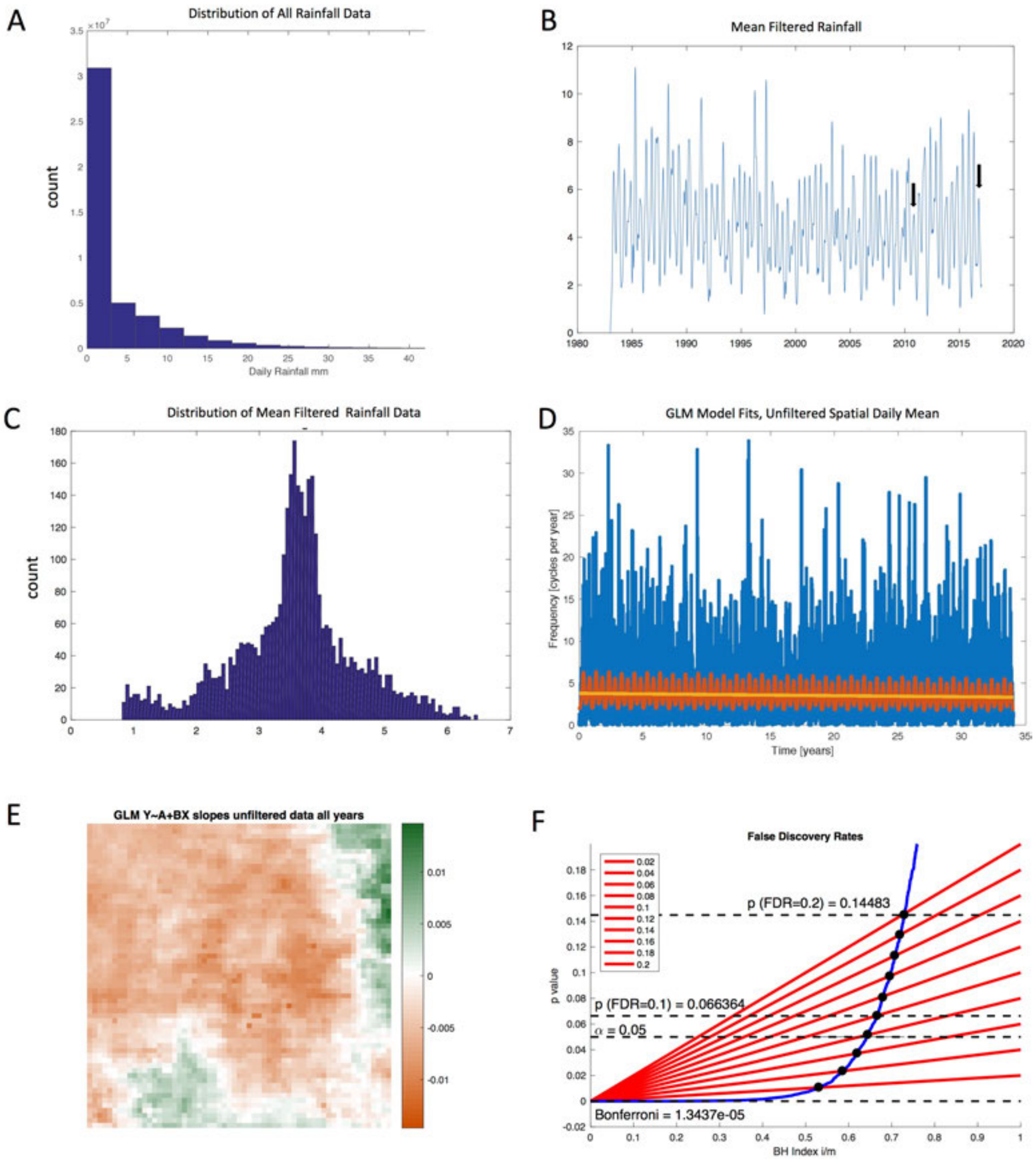

*Figure 2.* Spatially averaged rainfall demonstrates a decrease over the 34 year record. The nature of the data with and without spatial averaging is shown, and the origin of the average decrease in rainfall explained by the geographical distribution of decreases and increases in rainfall. A) Histogram of daily rainfall for all days and locations, truncated below 40 mm rain per day. B) Daily rainfall spatially averaged, and filtered with finite impulse response filter (order 100) for frequencies 0.05-6 cycles per year. Filter applied with zero phase filtering forwards and backwards. The twice-yearly rainy seasons in the East African Highlands are now plainly seen. Two drought anomalies from droughts in fall rainy season failures are indicated for 2010 and 2016 with black arrows. The averaged and filtered data set from B demonstrates a much less skewed distribution in C. D) Linear GLM fit, *log(μ)=A+BT,* shown by yellow line, and full model fitted with 0.25, 0.5, 1, 2, and 4 cycle per year frequencies shown by red line, superimposed on spatially averaged but unfiltered data in D. In E are shown the individual slopes of linear GLM regression to each grid point time series, with no averaging, on a scale illustrating the intensity of the negative (brown) or positive (green) slopes. F) False Discovery Rate (FDR) plot for the GLM fit p-values from the slope plot in E, for a family of FDR values. The intersections of the sorted p-values (blue) with the Benjamini-Hochberg coefficients (red lines) form the FDR thresholds employed in Figure 3.

**Figure 3**

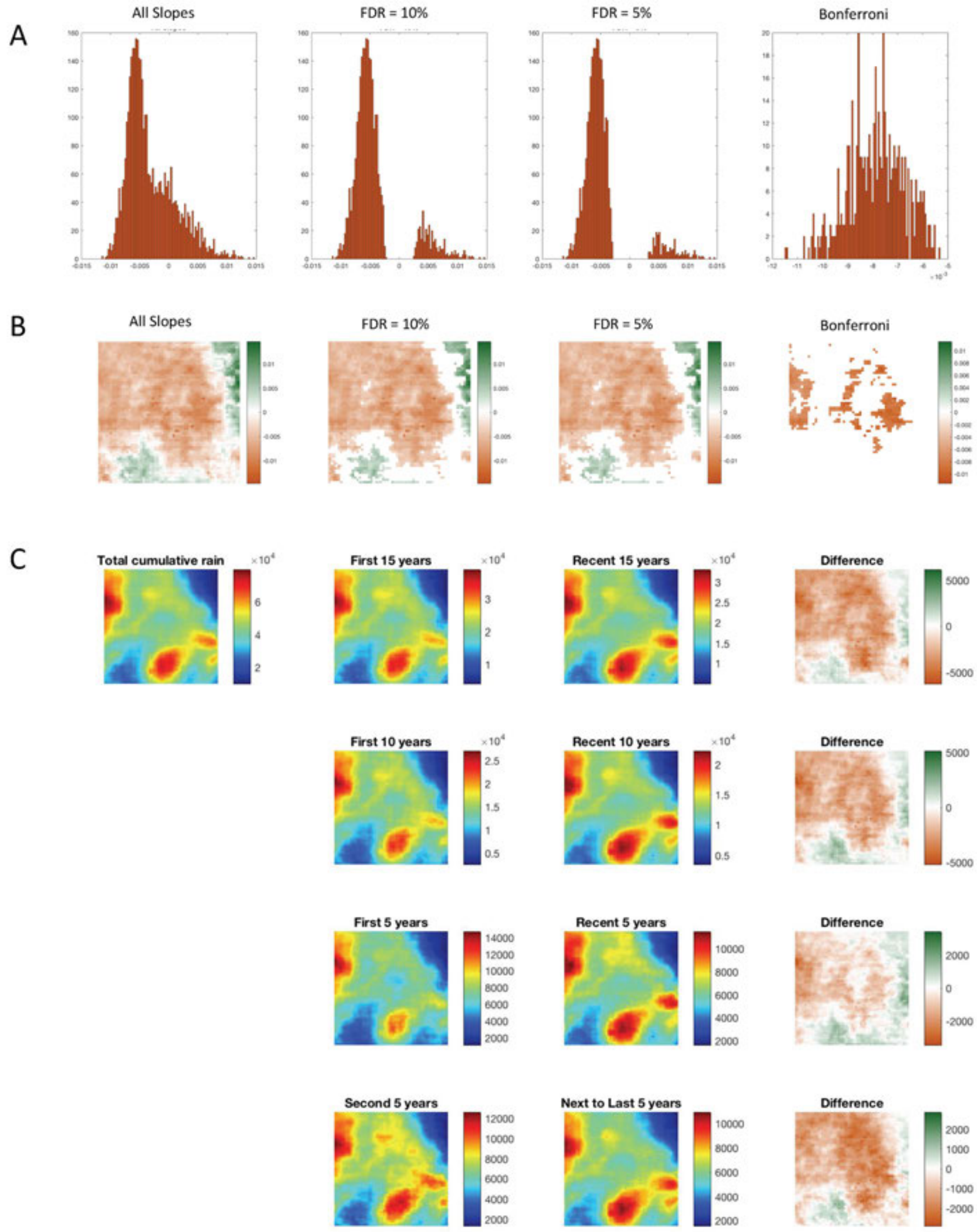

*Figure 3.* The decrease in rainfall is a robust finding. We first performed an exploration of false discovery rates, and then independently tested with cumulative rainfall differences. A) Histogram of the 3,721 GLM slopes from in Figure 2E. With progressively more conservative FDR rates, we see the progressive elimination of small magnitude slopes, culminating in the most conservative thresholding with Bonferroni ($\alpha/3721$). In B we map these distributions of slopes back onto the satellite grid. Note that the most significant slopes retain the coherent regions of central and western Uganda where the rainfall decreases the most over the historical record. To independently test these findings, we examine the difference maps of cumulative rainfall from the raw ARC2 data in C. Indeed, for the first and most recent 15, 10, and 5 year cumulative data, and the second and second-to-last 5 year periods, the difference maps retain the same rainfall pattern of decrease indicated from the GLM fits.

**Figure 4**

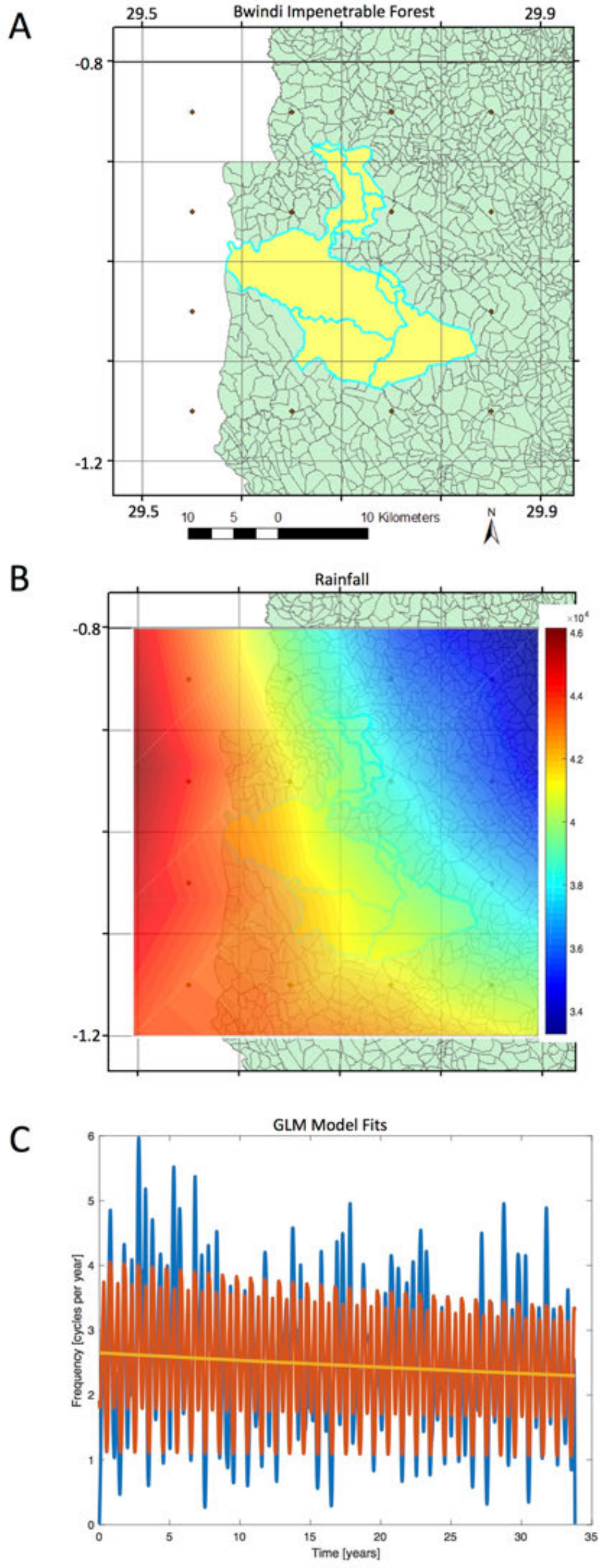

*Figure 4.* The decrease in rainfall is also reflected within the Bwindi Impenetrable Forest region. This is one of the last remaining habitats of the Mountain Gorilla. A) The Bwindi Impenetrable Forest in Uganda (yellow), covered by 16 of the satellite grids. Figure 11. B) Cumulative rainfall from 1983-2016 with interpolated shading over the Bwindi coordinates. C) Spatially averaged filtered time series of the Bwindi rainfall, overlain with the GLM models as in Figure 2. Note the substantial rainfall decrease over the record.

## Figure 5

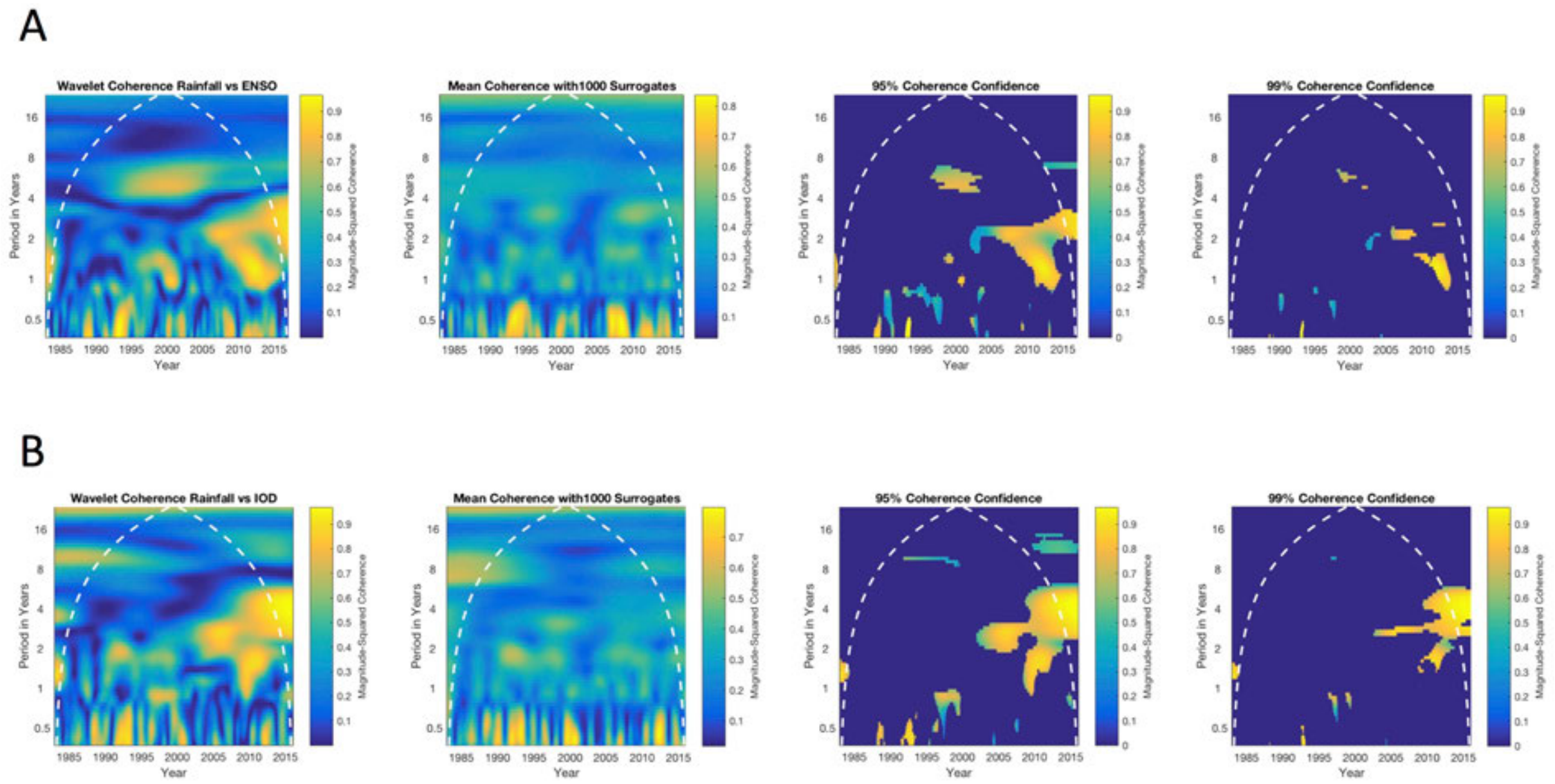

*Figure 5*. El Nino Southern Oscillation (ENSO) and Indian Ocean Dipole (IOD) relationship to spatially averaged rainfall. With rigorous statistical confidence bounds, there is a coherency between these indices and the Ugandan rainfall, but only in recent years, and only for relatively short periods from 1-4 years. A) Wavelet coherency between spatially averaged and filtered rainfall, with the ENSO index interpolated from monthly to daily values, and filtered identically as rainfall data. The second panel reflects the mean of 1000 surrogate coherency calculations, from which 95% (third panel) and 99% (fourth panel) confidence limits reveal only the regions that met these significance criteria. The cone of influence, delimiting where edge effects substantially confound the analysis, is indicated by the white dotted line. B) identical calculations for the IOD index. The most significant coherencies between ENSO and IOD occurs during the most recent 10 years, with periods from 1-4 years.